\documentclass[12pt]{article}

\newcommand{\beq}{\begin{equation}}
\newcommand{\beql}[1]{\begin{equation}\label{eq:#1}}
\newcommand{\eeq}{\end{equation}}
\newcommand{\be}{\begin{equation}}
\newcommand{\ee}{\end{equation}}
\newcommand{\beqn}{\begin{eqnarray}}
\newcommand{\eeqn}{\end{eqnarray}}
\newcommand{\bea}{\begin{eqnarray}}
\newcommand{\eea}{\end{eqnarray}}

\DeclareFixedFont{\xiiss}{OT1}{cmss}{m}{n}{12}
\DeclareFixedFont{\ixss}{OT1}{cmss}{m}{n}{9}
\DeclareFixedFont{\cmrnine}{OT1}{cmr}{m}{n}{9}

\newcommand{\CC}{\hbox{\xiiss C\kern-.4emI}}
\newcommand{\RR}{\hbox{\xiiss R\kern-.45emI}}
\newcommand{\ZZ}{\hbox{\xiiss Z\kern-.4emZ}}
\newcommand{\CCs}{\hbox{\ixss C\kern-.4emI}}
\newcommand{\ZZs}{\hbox{\ixss Z\kern-.4emZ}}
\newcommand{\pa}{\partial}

\newcommand{\pasl}{\pa\kern-.55em /}
\newcommand{\Dsl}{D\kern-.65em /}

\def\tx{{\tilde{x}}}
\def\vx{{\bf x}}
\def\tp{{\tilde{\psi}}}
\def\tal{{\tilde{\alpha}}}
\def\href#1#2{#2}
\def\dr{\rangle\!\rangle}

\def\coherentr#1#2#3{|D{#1};{#2};{#3}\rangle}

\def\bz{{\bf Z}}

\begin{document}
\begin{titlepage}
\title{ 
        \begin{flushright}
        \begin{small}
        RU-99-16\\
        hep-th/9905024\\
        \end{small}
        \end{flushright}
        \vspace{1.cm}
 D-branes on Asymmetric Orbifolds}
\author{
Ilka Brunner\thanks{e-mail: \tt ibrunner@physics.rutgers.edu}, 
Arvind Rajaraman\thanks{e-mail: \tt arvindra@physics.rutgers.edu}
\ and
Moshe Rozali\thanks{e-mail: \tt rozali@physics.rutgers.edu}\\
\\
        \small\it Department of Physics and Astronomy\\
        \small\it Rutgers University\\
        \small\it Piscataway, NJ 08855
}

\maketitle

\begin{abstract}
We construct D-brane states on an asymmetric orbifold
of type IIA on a four-torus, which is modded out by T-duality.
We find explicit boundary states charged under the twisted 
sector gauge fields. Unlike other cases, the boundary
states involve an explicit dependence on the twist
fields. The D-brane spectrum is consistent
with the  model being equivalent to type IIA
on a four-torus.
\end{abstract}

\end{titlepage}

%%%%%%%%%%%%%%%%%%%%%%%%%%%%%%%%%%%%%%%%%%%%%%%%%%%%%%%%%%%%%%%%%%%%%%%%%%%%%

\section{Introduction}

Asymmetric orbifolds are string vacua where the orbifold
action acts differently on the left- and right-movers \cite{asymm}.
Since the left- and right-moving bosons on the worldsheet
see different target spaces, these vacua usually do
not have a simple geometrical interpretation.
Such vacua are
potentially interesting for phenomenology because
they typically have very few moduli. In certain cases,
all moduli except the dilaton are projected out \cite{islands}.

Since these spaces have no geometrical interpretation,
it is difficult to ascertain where they lie in the web of
string dualities. In particular, it is unknown whether
all such orbifolds are smoothly connected to 
large volume  Calabi-Yau spaces. As a consequence, an
M-theoretic interpretation of these vacua is missing.

Based on the experience with other string dualities,
it is natural to guess that the clue to understanding the
non-perturbative structure of these orbifolds is
the understanding of  their D-brane spectra.
D-branes have been shown to probe sub-stringy structure,
and in other cases (e.g. symmetric orbifolds), the metric
on the D-brane worldvolume reproduces expected properties
of the underlying geometry \cite{Douglas}. It is of great interest,
therefore,
to analyze the moduli space metric in cases where no underlying
geometric structure is known, as in the case of asymmetric
orbifolds.

Furthermore, D-branes control the strong coupling behaviour
of a theory in many cases, so the knowledge of the D-brane spectra
constrains possible dual models, and may also suggest guesses for the
dual theory.

Asymmetric orbifolds have also arisen in a recent attempt
to construct non-supersymmetric vacua with zero
cosmological constant \cite{kks}. It was shown  that in a particular
model the cosmological constant vanished to two loops and
it was conjectured to vanish to all orders in perturbation
theory. Non-perturbative contributions then become important,
and in \cite{Harvey} an example was analyzed where duality arguments
suggested the existence of a non-perturbative contribution to
the cosmological constant. It is of interest to see if direct
computations using D-branes in this model support this result.
So again we are led to the study of D-brane spectra.

From a technical point of view, constructing D-branes
in asymmetric spaces is qualitatively different from
the symmetric case. D-branes are defined as endpoints of
strings, so at the boundary of the string worldsheet one
has to provide a boundary condition relating the left
moving fields  to the right-moving fields.
However, in asymmetric spaces, there is no obvious way to
write such a condition. Heuristically, this is why not
all string theories have D-branes.

 The  question studied in this paper is that
 of the existence (and construction)
of ``twisted'' D-branes,
i.e. D-branes that couple to R-R fields coming from a twisted sector.
 An untwisted brane can be constructed by summing all images 
 of a boundary state under the orbifold action \cite{hkms}.  However,  these
states are,
at least in the example studied here, not the minimal D-brane states. For
 example they preserve only a quarter of the supersymmetries.

We shall  construct D-branes in a special example,
where we can avoid many of the more difficult problems.
This is the case of type II string theory on an orbifold of the 
 four-torus, where the orbifold
action is that of an overall T-duality \cite{ds}. This turns out
to have a second description in terms of type II
string theory on $T^4$, where the D-brane spectrum is known.
In this case, using the equivalent description, 
we are able to construct the  D-branes explicitly.

We find that the D-branes indeed arise from twisted sectors
of the asymmetric orbifold. We describe these states explicitly
as boundary states \cite{torus, ishi}. Using
this formalism, we show that these D-branes  behave exactly
as expected of branes on a four-torus. In particular they are BPS
states, preserving
half of the supersymmetry. Furthermore, the
D-branes from the untwisted sector are shown to be
combinations of twisted D-branes. It is the twisted branes that play
the role of the elementary objects in the equivalent torus description.

This result leaves  important questions unanswered;
it is not known how to construct the twisted D-branes on general asymmetric
spaces. It, however, provides an existence proof of 
twisted D-branes, and emphasizes their importance.
% (and more generally of  D-branes on asymmetric orbifolds).
 Hopefully, the example here may also hint
at a more general construction. We hope to return to
these questions in the future.

The paper is organized as follows. We review in section 2 the general
formalism of boundary states as applied to toroidal backgrounds of
type II string theory. In section 3 we introduce the T-duality
orbifold, compute the closed string spectrum and  demonstrate the
spacetime symmetries of the models. These symmetries are used in
section 4 to construct the boundary state for a general BPS
 saturated D-brane in this model.

 While this work was in progress, we became aware of \cite{hkms},
 where related issues are discussed. 

\section{Boundary States on a  Torus}
Let us briefly review the construction of boundary states
\cite{torus, ishi} in  toroidal backgrounds. We denote by  $x^6,\ldots,x^9$
four   compactified directions, with radii $R_6,\ldots,R_9$. To avoid introducing
(super)ghosts, we will work  in the light-cone gauge as in \cite{gg}.
We 
consider Dirichlet boundary conditions localizing the D-branes at the
origin of all non-compact directions.

The problem at hand is to  construct the 
boundary states $ |B\rangle $
satisfying the boundary conditions
\be
(L_n-\tilde{L}_{-n})|B\rangle =0
\ee

This condition ensures that the resulting open string theory
is conformally invariant. Further conditions (called
Cardy's conditions \cite{cardy}) are necessary  to make the open
string sector
 a sensible boundary CFT  (e.g. degeneracies should be
integral).

In practice, the above condition is  hard to solve.
Instead one imposes (in the free field case) the more
restrictive condition \cite{ishi}
\bea
(\alpha^i_n-\tal^i_{-n}) |B\rangle=0\nonumber\\
(\alpha^\mu_n+\tal^\mu_{-n}) |B\rangle=0,
\eea
where $i=p+1,...,9$ labels the Dirichlet directions, and
$\mu$ labels the Neumann directions.
This manifestly implies (1), since
\bea
L_n=\sum_m{\left(\alpha^i_m\alpha^i_{n-m}
+\alpha^\mu_m\alpha^\mu_{n-m}
\right)}\nonumber \\
 \tilde{L}_{-n}=\sum_m{\left(\tal^i_{-m}\tal^i_{m-n}+
\tal^\mu_{-m}\tal^\mu_{m-n}\right)}
\eea
Note that the condition requires the $U(1)^4$ invariance of the torus,
or generally a chiral symmetry algebra larger than simply the Virasoro
algebra \cite{ishi}.

The condition (2) is solved, in the bosonic case, by the coherent state
\bea
|Dp;k\rangle =\exp\{\sum_{n>0}{1\over{n}}(\alpha^i_{-n}\tal_{-n}^i-\alpha_{-n}^\mu\tal_{-n}^
\mu)\}|{k}\rangle,
\eea
where $|{k}\rangle$ denotes the string ground state of momentum
$k^i$ in the $i$ directions and winding $m^\mu$ in the
$\mu$ directions.  This state is called an Ishibashi state.

To maintain worldsheet supersymmetry, we need to
impose conditions on the fermions as well.
In the NS-NS sector, the conditions are
\bea
(\psi_r^i-i\eta\tp_{-r}^i)\coherentr{p}{\eta}{k}_{_{NSNS}}=0 \nonumber\\
(\psi_r^\mu+i\eta\tp_{-r}^\mu)\coherentr{p}{\eta}{k}_{_{NSNS}}=0
\eea
where $\eta$ is a sign needed for GSO projection.

 The corresponding Ishibashi state is
\bea
\coherentr{p}{\eta}{k}_{_{NSNS}}&=&
\exp\{\sum_{n>0}{1\over{n}}(\alpha^i_{-n}\tal_{-n}^i-\alpha_{-n}^\mu\tal_{-n}^
\mu)\}|{k}\rangle
\nonumber \\
&&\otimes
\exp\{i\eta\sum_{r>0}\psi^i_{-r}\tp^i_{-r}-\psi^\mu_{-r}\tp^\mu_{-r}\}|F\rangle_
{_{NSNS}}
\eea
where $|F\rangle_{_{NSNS}}$ is the NSNS fermionic Fock
vacuum.

To GSO project, we note that $(-1)^F$ and $(-1)^{\tilde{F}}$ act as
\bea
(-1)^F|Dp;\eta;k\rangle_{_{NSNS}}=-|Dp;-\eta;k\rangle_{_{NSNS}}
\nonumber \\
(-1)^{\tilde F}|Dp;\eta;k\rangle_{_{NSNS}}=-|Dp;-\eta;k\rangle_{_{
NSNS}}
\eea
and hence the GSO projected state is
\bea
|Dp;k\rangle_{_{NSNS}}={1\over{\sqrt{2}}}\left(|Dp;+;k\rangle_{_{NSNS}}-|Dp
;-;k\rangle_{_{NSNS}}\right)
\eea

We now  perform the same analysis in the RR sector. The
boundary conditions are then
\bea
\label{bc}
(\psi_r^i-i\eta\tp_{-r}^i)\coherentr{p}{\eta}{k}_{_{RR}}&=&0 \nonumber\\
(\psi_r^\mu+i\eta\tp_{-r}^\mu)\coherentr{p}{\eta}{k}_{_{RR}}&=& 0
\eea
with the solution
\bea
\coherentr{p}{\eta}{k}_{_{RR}}&=&
\exp\{\sum_{n>0}{1\over{n}}(\alpha^i_{-n}\tal_{-n}^i-\alpha_{-n}^\mu\tal_{-n}^
\mu)\}|{k}\rangle\nonumber \\
&&\otimes
\exp\{i\eta\sum_{r>0}\psi^i_{-r}\tp^i_{-r}-\psi^\mu_{-r}\tp^\mu_{-r}\}|F_\eta
\rangle_
{_{RR}},
\eea
where $|F_\eta\rangle_{_{RR}}$ is an appropriately chosen  R-R  
vacuum. The Fock
vacuum of the R-R sector carries left- and right spinor indices, and
as a result of the condition (\ref{bc}) for the fermionic zero modes,
is constrained to satisfy \cite{berg,sen,new}:
\bea
(\psi_0^i-i\eta\tp_0^i)|F_\eta\rangle_{_{RR}} =0 \nonumber \\
(\psi_0^\mu+i\eta\tp_0^\mu)|F_\eta\rangle_{_{RR}} =0 
\eea

The action of $(-1)^F$ and $(-1)^{\tilde{F}}$ is now
\bea
(-1)^F|Dp;\eta;k\rangle_{_{RR}}&=&(-1)^{7-p}|Dp;-\eta;k\rangle_{_{RR}}
\nonumber \\
(-1)^{\tilde F}|Dp;\eta;k\rangle_{_{RR}}&=&|Dp;-\eta;k\rangle_{_{RR}}
\eea
Therefore the GSO projected state in the R-R sector is
\bea
|Dp;k\rangle_{_{RR}}={1\over{\sqrt{2}}}\left(|Dp;+;k\rangle_{_{RR}}+|Dp;-
;k\rangle_{_{RR}}\right)
\eea

To make the state spacetime supersymmetric, we need to combine
the NS-NS and R-R sectors appropriately. The result turns out
to be \cite{sen,irc}
\bea
|Dp;k\rangle={1\over{\sqrt{2}}}\left(|Dp;k\rangle_{_{NSNS}}\pm4
i|Dp;k\rangle_{_{RR}}\right)
\eea
where the $+,-$ signs refer to branes and anti-branes respectively.

We still need to satisfy Cardy's conditions. This is achieved
by taking a linear combination of the above states.
In effect, this is a Fourier transform from
momentum basis to position basis. The result is \cite{irc}
\bea
|Dp;\vx\rangle=\int\prod_{l=0}^5dk_l|Dp;k_l\dr_D
\times\prod_{j=6}^{6+p}{1\over{\sqrt{2R_j}}}\sum_{n_j\in\bz}e^{-in_jx_j\over R_
j}|Dp;n_j\dr_D\nonumber\\
\prod_{\mu=7+p}^{9}{\sqrt{R_\mu}}\sum_{m_\mu\in\bz}e^{-i2R_\mu m_\mu\tx_\mu}|Dp;m
_\mu\dr_N
\eea
where $n_j,m_\mu$ are the quantized momenta and winding in the compact
directions. 

The result is a D-brane state localized in the origin of the noncompact
 dimensions. The generalization to more general states is straightforward.

\section {The T-Duality Orbifold}

  We now discuss the example of type IIA theory compactified on an
asymmetric
orbifold of $T^4$. The orbifold group is chosen to be $Z_2$, generated
by a reflection of all left moving oscillators:
\bea
\alpha^i_n \rightarrow - \alpha^i_n \qquad \tal^i_n \rightarrow \tal^i_n
\nonumber \\
\psi^i_r \rightarrow -\psi^i_r \qquad \tilde{\psi}^i_r \rightarrow
\tilde{\psi}^i_r \\
|F\rangle_{RR} \rightarrow \Gamma^1 \Gamma^2\Gamma^3
\Gamma^4 |F\rangle_{RR} \nonumber
\eea
where $i=1,2,3,4$.

This 
action is an overall
T-duality, therefore we refer to this model as the T-duality
orbifold. This action is only a symmetry of the torus at an $SO(8)$
point\footnote{ To be precise, we must take the torus at
the $SO(8)$ point, rather than the ${SU(2)}^4$ point,   so that the
 zero mode contributions level match 
\cite{erler}. We thank R. Blumenhagen and E. Silverstein for
correspondence on this point.}. We are interested in computing the massless spectrum of this
model, using lightcone quantization of the RNS string. 

In order for the model to be consistent (modular invariant), one has
to check level matching. For an orbifold element of order $n$, one
has to satisfy \cite{vafa}:
\be
E^L-E^R = 0 \ \mbox{mod}\  \frac{1}{n}
\ee
 where $E^L, E^R$ are the ground state energies of the left- and right-
 movers
respectively. This guarantees that the physical states, which have to
 be invariant under the orbifold action, are level matched
 \cite{polchinski}.

Here the only non-trivial check is for the $Z_2$ generator. The ground
state
energies for the left movers include contributions from the twisted
bosons and fermions, as well as from the non-compact directions, and are
found to be:
\bea
E^L_R= 0 \nonumber\\
E^L_{NS} = 0
\eea

The right movers have the standard ground state energies:
\bea
E^R_R &=&0 \nonumber\\
E^R_{NS} &=& - \frac{1}{2}
\eea
therefore level matching is satisfied in all 4 sectors of the RNS
string.

The massless spectrum of the orbifold is identical to that of type II
on a 4-torus, making it natural to conjecture that these two models
are in fact equivalent, including their D-brane spectrum. To study the
mapping between these two vacua  we concentrate on the correspondence
between spacetime supersymmetry generators, as well as vector fields
corresponding to the isometries of the 4-torus.

 In the untwisted sector of the orbifold, one finds that 2 out of the 8
gravitini are projected out, leaving 6 gravitini (24 real components) 
in 6 dimensions. Specifically, in the R-NS sector, half of the
components
of the left-moving spin operator are not invariant under the orbifold
action, and are projected out. We denote the surviving left moving
spin operator by $S_\alpha$. It is defined by:
\be
\Gamma^1 \Gamma^2\Gamma^3
\Gamma^4 S_\alpha = S_\alpha
\ee
  Here $\alpha= 1,2$ is an $SU(2)$ index, which is denoted by $SU(2)_I$.

On the other hand, two supersymmetry generators return in the twisted
sector.
The R-NS sector of the twisted sector contains one gravitino, made
entirely
from non-compact bosons and fermions. This is because the internal
fermions in the  twisted Ramond sector have no zero modes. The 
degeneracy of the twisted
sector
ground state is computed as in \cite{asymm}, and is found to be
2.  Altogether the model
has the maximal supersymmetry in six dimensions.

 The study of the spacetime vectors is related to the isometries of
 the 4-torus, and is important in constructing D-brane states.

On the orbifold model, 4 vectors arise in the untwisted sector
\be
V_L = \psi^{\mu}_{-1/2}  \tilde{\psi^r}_{-1/2}|F\rangle_{NSNS} \\
\ee
where the directions of the 4-torus in  the orbifold model are denoted
by the index $r= 1,...,4$.

The additional 4 vectors arise in the twisted sector as follows. The
twisted sector vacuum for the left movers is generated by 
 bosonic twist operators
acting on the usual  NS vacuum. There are two such twist
operators, $\tau_\beta$,  as the ground state degeneracy of the
 twisted sector is 2.  Here $\beta$ is also an $SU(2)$ index; we
denote this $SU(2)$ by $SU(2)_T$.

  The
bosonic twist operators $\tau_\beta$ have dimension $\frac {1}{4}$. In addition
there is a fermionic twist operator, which in the $Z_2$ case reduces
to the usual spin operator.
   There are 2  spin operators, $S_\alpha$, that 
are invariant
under the orbifold action. They have dimension  $\frac {1}{4}$ as
well, so the total twist operator carries dimension $\frac{1}{2}$.
 In summary we find that the left moving ground state has 
degeneracy 4, which we
denote by an $SO(4)= SU(2)_I \times SU(2)_T$ vector index $r$.

 These twisted ground states are
generated from  the NS vacuum by 4 operators of dimension
$\frac{1}{2}$,
which are denoted for later convenience as $J^r$. It is easy to show
that
$J^r$ are GSO odd. Using the mapping of
$SU(2)_I \times SU(2)_T$ into $SO(4)$ we can write:
\be
\label{curr}
J^r = (\Gamma^r)_{\alpha \beta} S^\alpha \tau^\beta
\ee

The additional 
4 massless vectors are then:

\be
V_R  = 
\tilde{\psi}^{\mu}_{-1/2} J^r|F\rangle_{NSNS}
\ee
 Note that they are GSO even as required. 

  We wish to compare these states with their counterparts
in the torus description. Type II
 string theory on the 4-torus has 8 massless vectors in the NS-NS
 sector, 
related to momentum and 
 winding modes
on the 4-torus. The corresponding states are:
\bea
V_L &=& \psi^{\mu}_{-1/2}  \tilde{\psi^a}_{-1/2}|F\rangle_{NSNS} \nonumber \\
V_R &=& \tilde{\psi}^{\mu}_{-1/2} \psi^a_{-1/2} |F\rangle_{NSNS}
\eea
where $\psi^a,\tilde{\psi^a} $ are fermions in the 4-torus 
directions $(a= 1,...,4)$, and
$\psi^{\mu},\tilde{\psi}^{\mu}$
are fermions in the noncompact directions.

It 
is then natural to identify the new torus fields as:
\bea
\tilde{\psi^r} \rightarrow \tilde{\psi^a} \nonumber\\
J^r \rightarrow \psi^a 
\eea

Similarly in the left and right-moving Ramond sectors, the ground
states
carry spinor indices of SO(4). The spinor indices
of the non-compact directions are suppressed. Denote, as above, the
spinor index which is invariant under reflection by $\alpha$, and the
other spinor index by $\tilde{\alpha}$. The ground state degeneracy in
the twisted sector is parametrized by the index $\beta$. 

With these notations, the right moving spinors of the new torus can be
identified as $\tilde{S}_\alpha$ and $\tilde{S}_{\tilde{\alpha}}$. The 
left-moving spinors are $S_\alpha$ and $\tau_\beta$. 

 Having identified the torus symmetries in the asymmetric orbifold
 language, we are ready to use those symmetries to write the D-brane
 boundary
states.

\section{Boundary States}

 In the T-duality orbifold, it is possible to construct ``untwisted''
  branes by starting with any brane on the torus, and adding its images
under the $Z_2$ action. For example a 0-brane with any location on the
 torus
is T-dual to a  4-brane wrapped on the torus, with some Wilson lines on
 its worldvolume.  The $Z_2$ invariant boundary state is simply the
 sum of these two states.  It is charged only with respect to
 untwisted sector fields, hence the name  ``untwisted''.

 The interpretation of this state in the new torus description is not trivial.
 The locations and Wilson lines are separate moduli of the
 two components in the boundary state. As the state is required to be
 $Z_2$
invariant, its moduli space is $T^4$. It is therefore
natural to interpret this untwisted state as a single  D-brane state
on the new torus. However, this is incorrect since the untwisted
brane preserves only a quarter of the supersymmetry generators. In the
 equivalent torus description such states are made from two types of
 D-branes, and have a larger moduli space.

Therefore, similar to the closed string spectrum, the
untwisted branes give only a subset of the expected D-brane states on
 the new torus. In particular,  the naive moduli space of an untwisted
brane is incorrect unless one allows more general twisted branes.

In order to construct the general  D-brane state we are therefore 
forced to consider 
twisted branes, that is, branes that carry some twisted R-R
charge. For a 
general asymmetric orbifold this is
 a difficult task. Constructing
boundary states is usually done by imposing not just conformal
invariance, but a more restrictive invariance under some chiral
algebra.  By construction, the chiral algebras for the left or right movers
are different  in most twist sectors. It is then impossible to 
use the usual ansatz to construct boundary states in each twist sector
separately. 

However, in the case considered in this paper, one can use the symmetries
identified in the last section to construct boundary states.  We 
start by constructing the NS-NS component of the boundary states.

 The right movers of the new torus were identified as the fields
$\tilde{\psi^r}$ of dimension (0, $\frac{1}{2}$) and their partners
$\tilde{\partial}X^r$
of dimension (0, 1). The left movers are the twist fields $J^r$ of
dimension ($\frac{1}{2}$, 0), and their partners denoted as $T^r$, of 
dimension
(1, 0). We denote the modes of any operator $O$ by $O_n$, except the modes
of $\partial X$ and $\bar{\partial} X$, which are denoted  by $\alpha_n$ and
$\tilde{\alpha}_n$ respectively.

 It is now 
straightforward to write the Ishibashi states
for an arbitrary D-brane state. For example any D-brane with Dirichlet
boundary condition in the torus directions (i.e. an unwrapped brane)
has the 
factor:

\be
 |B_{\eta}\rangle_{NSNS} = \exp \left(\sum_n T^r_{-n} \tilde{\alpha}^r_{-n} +i
\eta\sum_n
J^r_{-n} \tilde{\psi}^r_{-n}\right) |F\rangle_{NSNS}
\ee

 This is multiplied by the part of the boundary state coming from the
 non-compact directions, which is given in section 2. One has to sum
 over the states$|B_\eta\rangle_{NSNS}$ in order to 
project into GSO invariant states, as one
 does in flat space.

 As explained in section 2, in order to preserve supersymmetry, one
 has to consider also
R-R boundary states. To find the correct combination of sectors to
generate a BPS state we are guided by the explicit form of the
spacetime supersymmetry operators.
 Intuitively,
 for the complete boundary state to be BPS
saturated, 
the R-R boundary states are 
to be related to the NS-NS boundary state by spacetime supersymmetry operators
(spectral flow).

In the present case the situation is similar.
The supersymmetry generators from the left-moving sector are generated by
the operators $S_\alpha$ and $\tau_\beta$ (the spin operators from the
non-compact directions are suppressed). Those two operators create
square root cuts in $J^r$, defined in (\ref{curr}).   Therefore we
have to consider
the sectors defined by the insertion of the supersymmetry generators at the
origin. Those sectors define unusual, mixed,  periodicity conditions for the
original bosons and fermions, but act simply on the new variables
$J^r$, $T^r$. We refer to these sectors loosely as the Ramond sectors
for the left movers.

In each such sector the fermionic currents $J^r$ are now integer
moded. The construction of the R-R boundary states is identical to
the one in section 2. One obtains :
\bea
|B_\eta\rangle_{RR}&&= \nonumber \\
&&\exp\left(\sum_{n>0}{1\over{n}}(T^i_{-n}\tal_{-n}^i)
+i\eta\sum_{r>0}J^i_{-r}\tp^i_{-r}\right)|F_\eta
\rangle_
{_{RR}}
\eea

We have suppressed the dependence on the noncompact directions. The 
vacua $|F_\eta
\rangle_
{_{RR}}$ are the vacua of the R-R sectors, where the left-mover
sector is a Ramond sector in the sense defined above.
 Those vacua 
 have to be chosen now to satisfy:
\bea
(J_0^i-i\eta\tp_0^i)|F_\eta\rangle_{_{RR}} =0 \nonumber \\
(J_0^\mu+i\eta\tp_0^\mu)|F_\eta\rangle_{_{RR}} =0 
\eea

 The complete boundary state can be written as:
\bea
\label{bs}
|B\rangle={1\over{\sqrt{2}}}\left(|B\rangle_{_{NSNS}}\pm4
i|B\rangle_{_{RR}}\right)
\eea

Furthermore, in order to obtain Neumann boundary
 conditions in any of torus directions one can modify the boundary
 state
as described in section 2. The GSO projection acts, as in type IIA,
 asymmetrically. Hence, only even dimensional branes survive. We thus
 obtain the complete spectrum of D-branes in type IIA on $T^4$.

As a final note we comment on the untwisted D-brane. Since some of the 
left-moving supersymmetry generators come from the twisted sector, it is clear
that
one cannot obtain a BPS state that breaks only half of the
supersymmetry when one uses only the untwisted fields. Given the general 
D-brane state above, equation (\ref{bs}), one
can recover any untwisted brane by an appropriate linear combination 
of those states. One then verifies the claim that the untwisted  branes are
two-object states which leave only a quarter of the supersymmetries unbroken.

\section{Acknowledgments}  

We thank O. Aharony,
T. Banks, R. Blumenhagen, M. Douglas, G. Moore,  J. Harvey and E. Silverstein
for useful conversations.

This work was supported in part by DOE grant DE-FG02-96ER40559.


\newpage
 
\begin{thebibliography}{99}
\small
\parskip=0pt plus 2pt
\bibitem{asymm}
K.S.~Narain, M.H.~Sarmadi and C.~Vafa,
``Asymmetric Orbifolds,"
Nucl. Phys. {\bf B288}, 551 (1987).

\bibitem{islands}
A.~Dabholkar and J.A.~Harvey,
``String islands,"
JHEP {\bf 02}, 006 (1999)
hep-th/9809122.

\bibitem{Douglas}
M.R.~Douglas and G.~Moore,
``D-branes, quivers, and ALE instantons,"
hep-th/9603167;\\
C.V.~Johnson and R.C.~Myers,
``Aspects of type IIB theory on ALE spaces,"
Phys. Rev. {\bf D55}, 6382 (1997)
hep-th/9610140;\\
M.R.~Douglas and B.R.~Greene,
``Metrics on D-brane orbifolds,"
Adv. Theor. Math. Phys. {\bf 1}, 184 (1998)
hep-th/9707214;\\
M.R.~Douglas, B.R.~Greene and D.R.~Morrison,
``Orbifold resolution by D-branes,"
Nucl. Phys. {\bf B506}, 84 (1997)
hep-th/9704151.


\bibitem{kks}
S.~Kachru, J.~Kumar and E.~Silverstein,
``Vacuum energy cancellation in a nonsupersymmetric string,"
Phys. Rev. {\bf D59}, 106004 (1999)
hep-th/9807076;\\
S.~Kachru and E.~Silverstein,
``Selfdual nonsupersymmetric type II string compactifications,"
JHEP {\bf 11}, 001 (1998)
hep-th/9808056;\\
S.~Kachru and E.~Silverstein,
``On vanishing two loop cosmological constants in nonsupersymmetric strings,"
JHEP {\bf 01}, 004 (1999)
hep-th/9810129.

\bibitem{Harvey}
J.A.~Harvey,
``String duality and nonsupersymmetric strings,"
Phys. Rev. {\bf D59}, 026002 (1999)
hep-th/9807213.

\bibitem{hkms}
J. Harvey, S. Kachru, G. Moore and E. Silverstein, to appear.

\bibitem{ds}
M.~Dine and E.~Silverstein,
``New M theory backgrounds with frozen moduli,"
hep-th/9712166.


\bibitem{torus}
C.G.~Callan, C.~Lovelace, C.R.~Nappi and S.A.~Yost,
``String loop corrections to beta functions,"
Nucl. Phys. {\bf B288}, 525 (1987); \\
J.~Polchinski and Y.~Cai,
``Consistency of open superstring theories,"
Nucl. Phys. {\bf B296}, 91 (1988).



\bibitem{ishi}
N.~Ishibashi,
``The boundary and crosscap states in conformal field theories,"
Mod. Phys. Lett. {\bf A4}, 251 (1989).




\bibitem{gg}
M.B.~Green and M.~Gutperle,
``Light cone supersymmetry and D-branes,"
Nucl. Phys. {\bf B476}, 484 (1996)
hep-th/9604091.

\bibitem{vafa}
C.~Vafa,
``Modular invariance and discrete torsion on orbifolds,"
Nucl. Phys. {\bf B273}, 592 (1986).



\bibitem{cardy}
J.L.~Cardy,
``Boundary conditions, fusion rules and the Verlinde formula,"
Nucl. Phys. {\bf B324}, 581 (1989).

%\bibitem{Recknagel:1997sb}
%A.~Recknagel and V.~Schomerus,
%``D-branes in Gepner models,"
%Nucl. Phys. {\bf B531}, 185 (1998)
%hep-th/9712186.

\bibitem{polchinski}
J. Polchinski, ''String Theory'', volume II, Cambridge University
Press (1998).




\bibitem{erler}
J.~Erler,
``Asymmetric orbifolds and higher level models,"
Nucl. Phys. {\bf B475}, 597 (1996)
hep-th/9602032; \\
R.~Blumenhagen and L.~Gorlich,
``Orientifolds of nonsupersymmetric asymmetric orbifolds,"
hep-th/9812158.

\bibitem{berg}
O.~Bergman and M.R.~Gaberdiel,
``A nonsupersymmetric open string theory and S duality,"
Nucl. Phys. {\bf B499}, 183 (1997)
hep-th/9701137.



\bibitem{new}
P.~Di Vecchia, M.~Frau, I.~Pesando, S.~Sciuto, A.~Lerda and R.~Russo,
``Classical p-branes from boundary state,"
Nucl. Phys. {\bf B507}, 259 (1997)
hep-th/9707068.

\bibitem{irc}
I. Brunner, R. Entin and C. R\"omelsberger, to appear.


\bibitem{sen}
A.~Sen,
``Stable non-BPS states in string theory,"
JHEP {\bf 06}, 007 (1998)
hep-th/9803194;\\
A.~Sen,
``Non-BPS states and branes in string theory,"
hep-th/9904207.

\end{thebibliography}
\end{document}